# SMEmail - A New Protocol for the Secure E-mail in Mobile Environments[†]

Mohsen Toorani
ResearcherID: A-9528-2009

*Abstract*- The electronic mail plays an unavoidable role in the humankind communications. With the great interest for the connection via mobile platforms, and the growing number of vulnerabilities and attacks, it is essential to provide suitable security solutions regarding the limitations of resource restricted platforms. Although some solutions such as PGP and S/MIME are currently available for the secure e-mail over the Internet, they are based on traditional public key cryptography that involves huge computational costs. In this paper, a new secure application-layer protocol, called SMEmail, is introduced that provides several security attributes such as confidentiality, integrity, authentication, non-repudiation, and forward secrecy of message confidentiality for the electronic mails. SMEmail offers an elliptic curve-based public key solution that uses public keys for the secure key establishment of a symmetric encryption, and is so suitable for the resource restricted platforms such as mobile phones.

## I. INTRODUCTION

Currently, the electronic mail (e-mail) plays an unavoidable role in the humankind communications. With the increasingly growing reliance on electronic mail in one hand, and the growing number of vulnerabilities and attacks on the other hand, there is an increasingly demand for the security solutions. Despite of the critical role of e-mail in the typical Internet user's life, e-mail as will be described later is not so secure.

The number of mobile subscribers has exceeded 3.33 billion people in the world [1] that is so greater than that of the Internet users. Mobile phones especially in the upper generations of cellular systems can connect to Internet, and the subscribers are then capable of sending and receiving e-mails. The processing capabilities of mobile phones are increasingly enhanced but it cannot compete with the processing capabilities of personal computers. There are also some additional security problems in the wireless media that are not the case in a wired system. Therefore, special secure protocols are required for the mobile platforms.

Two major solutions are available now that provide the end-to-end security for the e-mail: *Secure/Multipurpose Internet Mail Extension* (S/MIME) and *Pretty Good Privacy* (PGP), both of them are using the traditional public key cryptography [2] that involves huge computational costs and is not so suitable for the mobile devices that are dealing with processing and power limitations.

In this paper, a new protocol for the secure e-mail, called SMEmail, is introduced that provides several security attributes such as confidentiality, integrity, authentication, non-repudiation, and forward secrecy of message confidentiality. Since it deploys elliptic curves and the concept of signcryption, it decreases the computational costs and communication overheads, and it is so suitable for the mobile devices. This paper is organized as follows. The security of e-mail is briefly evaluated in Section 2. Section 3 is devoted to description of our proposed protocol (SMEmail), Section 4 takes a glimpse at its security attributes, and Section 5 provides the conclusions.

## II. THE SECURITY OF E-MAIL

Data Confidentiality, Authentication, Integrity, Non-repudiation, Access control, and Availability are the most important security services in the security criteria that should be taken into account in secure applications and systems [2]. However, there is no provision for such security services in the traditional e-mail protocols. E-mail is vulnerable to both passive and active attacks. Passive threats include *Release of message contents*, and *Traffic analysis* while active threats include *Modification of message contents*, *Masquerade*, *Replay*, and *Denial of Service* (DoS) [2]. Actually, all the mentioned threats are applicable to the traditional e-mail protocols [3, 4]:

- *Disclosure of Information*: Most of e-mails are currently transmitted in the clear (not encrypted). By means of some available tools, persons other than the designated recipients can read the e-mail contents.
- *Traffic analysis*: It is believed that some countries are routinely monitoring e-mail messages as part of their *Echelon* project. This is not just for counter-terrorism reasons but also to facilitate combat against industrial espionage and to carry out political eavesdropping [4]. However, it is not devoted to the national agencies since there is a thriving business in providing commercial and criminal elements with the information within e-mails.

- *Modification of messages*: E-mail contents can be modified during transport or storage. Here, the *man-in-the-middle* attack does not necessarily require the control of gateway since an attacker that resides on the same *Local Area Network* (LAN), can use an *Address Resolution Protocol* (ARP) spoofing tool such as "ettercap" to intercept or modify all the e-mail packets going to and from the mail server or gateway.
- *Masquerade*: It is possible to send a message in the name of another person or organization.
- *Replay of previous messages*: Previous messages may be resent to other recipients. This may lead to loss, confusion, or damage to the reputation of an individual or organization. It can cause some damage if e-mail is used for certain applications such as funds transferring, registration, and reservation.
- *Spoofing*: False messages may be inserted into mail system of another user. It can be accomplished from within a LAN, or from an external environment using Trojan horses.
- *Denial of Service:* It can put a mail system out of order by overloading it with mail shots. It can be carried out using Trojan horses or viruses sent to users within the contents of e-mails. It is also possible to block the user accounts by repeatedly entering wrong passwords in the login.

S/MIME and PGP as two major solutions that were developed for making the electronic mails secure, use public key cryptography for providing the same security services at the application layer. The main difference is that the former was intended as an industry standard for commercial and organization use while the latter is a choice for personal e-mail security. PGP tried to be independent of any governmental or standards organization. PGP uses a trust model instead of using the conventional digital certificates while S/MIME uses the well-tried X.509v3 certificates. S/MIME is a security enhancement to the MIME Internet e-mail format standard that secures a MIME entity with signature, encryption, or both of them [5]. Both PGP and S/MIME are supposed to be run on personal computers that take advantages of adequate processing capabilities. Their recommended approaches are exponentiation-based, and they follow the so-called "signature-then-encryption" scheme.

Currently, the *Elliptic Curve Cryptography* (ECC) has revolutionized the arena of security establishment. The EC-based solutions are usually based on difficulty of solving the *Elliptic Curve Discrete Logarithm Problem* (ECDLP) and factorization in elliptic curves [6]. The EC-based systems can attain to a desired security level with significantly smaller keys than those of required by their exponential-based counterparts. As an example, it is believed that a 160-bit key in an EC-based system provides the same level of security as that of a 1024-bit key in a RSA-based system [7]. This creates great efficiencies in key storage, certificate size, memory usage, and required processing so it enhances the speed and leads to efficient use of power, bandwidth, and storage that are basic limitations of resource constrained devices. The EC-based approaches can have a computationally advantage over their exponential-based counterparts, especially for the resource restricted platforms. As a practical example, the Infineon's SLE 66CUX640P security controller that has usages in USB applications, executes an elliptic curve point multiplication (with a 160-bits modulus) in an average time of 83ms while it carries out a modular exponentiation (with a 1024-bits modulus) in an average time of 220ms, both at the clock frequency of 15MHz [7].

In the traditional signature-then-encryption schemes, the plaintext is digitally signed, and then the plaintext plus signature are encrypted. Such summation has two problems: Low efficiency and high costs of such summation, and the case that any arbitrary scheme cannot guarantee the security or may not provide certain security attributes such as forward secrecy. Recently, the signcryption is proposed as a suitable alternative for the traditional signature-then-encryption schemes. The signcryption is a relatively new cryptographic technique that is supposed to fulfill the functionalities of digital signature and encryption in a single logical step, and can effectively decrease the computational costs and communication overheads in comparison with the traditional signature-then-encryption schemes. As a typical example, an EC-based signcryption scheme can save 58% of computational costs and 40% of communication overheads when it is compared with the traditional EC-based signature-then-encryption schemes [8]. Several signcryption schemes are proposed over the years, each of them offering different level of security services and computational costs. Our proposed protocol (SMEmail) that will be described in the next section uses a variant of a signcryption scheme that was initially developed in [9].

### III. THE PROPOSED SCHEME (SMEMAIL)

A brief description of our proposed protocol (SMEmail) is provided in this section. SMEmail is an application-layer protocol that provides an end-to-end security for electronic mails in mobile environments. The end-to-end security in the cellular systems can be generally provided by exploiting the processing capabilities of one or some of the following items [10]:

1) The *Mobile Equipment* (ME) using programming languages,
2) The SIM card using *SIM Application Toolkit* (SAT),
3) An additional smart card, e.g. JavaCard,
4) A crypto-processor that is embedded in the ME,
5) A portable PC connected to the ME.

SMEmail is based on the first solution, and suggests using J2ME (Java 2 Mobile Edition) as a programming platform that is designed for resource restricted devices such as mobile phones and *Personal Digital Assistants* (PDA), and is supported by most of currently available mobile phones. Consequently, SMEmail is implemented as a MIDlet that will be installed on the mobile phones of subscribers.

SMEmail is designed and adapted for the certain restrictions and specifications of mobile phones. It consists of initialization phase that is accomplished once at the initialization of the system, and message exchange phase in which the participants exchange their secured e-mail contents. Since SMEmail deploys *Public Key Cryptography* (PKC), there is a revocation phase that follows the typical rules of *Public Key Infrastructures* (PKI) but it is not considered in this paper.

*A. Initialization*

The initialization phase of SMEmail includes:
1) Selecting the domain parameters.
2) Installing the software application on ME.
3) Generating the public/private keys, and issuing a certificate for the public key of each user.

The domain parameters consist of a suitably selected elliptic curve $E$ defined over the finite field $F_q$ with the Weierstrass equation of the form $y^2 = x^3 + ax + b$, and a base point $G \in E(F_q)$ in which $q$ is a large prime number. In order to make the elliptic curve non-singular, $a, b \in F_q$ should satisfy $4a^3 + 27b^2 \neq 0 (\bmod q)$. To guard against small subgroup attacks [6], the point $G$ should be of prime order $n$ ($nG = O$ where $O$ denotes the point of elliptic curve at infinity), and we should have $n > 4\sqrt{q}$. To protect against other known attacks on special classes of elliptic curves, $n$ should not divide $q^i - 1$ for all $1 \leq i \leq f$ ($f = 20$ suffices in practice), $n \neq q$ should be satisfied, and the curve should be non-supersingular [6]. To guarantee the intractability of ECDLP to Pollard-rho and Pohlig-Hellman algorithms [11], $n$ should also satisfy $n > 2^{160}$. NIST SP 800-56A [12] has also specified the minimum bit lengths of elliptic curve's domain parameters for different required level of security.

As SMEmail uses a block cipher in its core for encrypting the e-mail contents, it is recommended to select a strong and efficient block cipher such as AES as a predefined algorithm for the symmetric encryption. However, it is also feasible to provide a list of block ciphers and let the user choose the suitable algorithm, as it is accomplished in PGP, but this is not recommended for a resource restricted device since it can increase the code size of application.

The private key of user $U$ is a randomly selected integer $SK_U \in_R [1, n-1]$, and the corresponding public key is generated as $PK_U = SK_U G$. The process of public/private key generation in SMEmail is depicted in Fig. 1. The CA server issues a certificate $Cert_U$ for the public key of each user. The certificates contain strings of information that uniquely identify users and bind their identities to their public keys. The X.509v3 certificates are a popular type of certificates that are extensively used, and they are also used for the SMEmail. Since e-mail addresses are unique in the world, they will be regarded as the unique identifier of each user ($ID_U$). The unique identifier, the public key, and the corresponding certificate of each user are also stored in an LDAP (Lightweight Directory Access Protocol) directory, as it is depicted in Fig. 1.

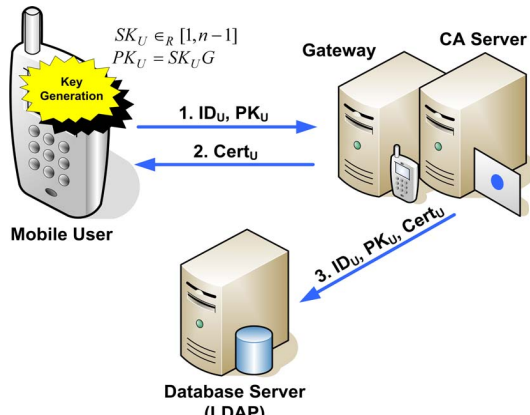

Figure 1. Public/private key generation in SMEmail

Since public/private keys are to be generated at end-entities, several precautions should be taken into account to thwart the potential threats. The CA should verify that each entity really possesses the corresponding private key of its claimed public key. This may be accomplished by a zero-knowledge technique. To thwart the *invalid-curve* attacks, the CA server should also check the validity of public keys. The public key of user $U$, $PK_U = (x, y)$ is valid if all the following conditions are simultaneously satisfied [13]:

(a) $U_U \neq O$.

(b) $x$ and $y$ should have the proper format of $F_q$ elements.

(c) $PK_U$ should satisfy the defining equation of $E$.

It is assumed that the participants have access to an authentic copy of the *CA*'s public key, in order to use it for the certificate validation. According to GSM 11.11 standard, the trusted keys and certificates (e.g. CA's) are stored in 4FXX files. The key information can be stored in file 4F50. The private key can be encrypted using a user's password or PIN (Personal Identification Number) and be stored in an elementary file of the SIM. The public keys can be stored in a transparent elementary file. The hash value of private key that may be used for validation of the encrypting password can also be stored in the SIM.

The application of SMEmail can be manually installed on the mobile phone if the mobile phone is capable of supporting such an application. It is also feasible to download and install the application via an *Over-The-Air* (OTA) server, as it is depicted in Fig. 2, where the *Application Management Software* (AMS) takes care of downloading and installing application on the device. A complete view of key generation and application installation using the OTA server is depicted in Fig. 2.

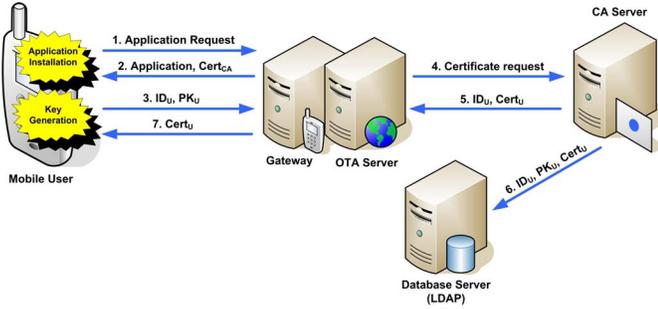

Figure 2. The key generation and application installation in SMEmail when the application installation is accomplished via an OTA server

*B. Message Exchange*

SMEmail is a certificate-based protocol so it is necessary for both of participants to perform the certificate validation for the certificate of the other party and use it for verifying the corresponding public key. The process of certificate validation includes [11]:

(a) Verifying the integrity and authenticity of the certificate by verifying the CA's signature on the certificate.
(b) Verifying that the certificate is not expired.
(c) Verifying that the certificate is not revoked.

The revocation status can be examined using either of *Certificate Revocation List* (CRL), Delta CRL, or *Online Certificate Status Protocol* (OCSP) [11]. Since CRL is too large for the limited memory capacity of mobile devices, SMEmail uses OCSP [14] for checking the revocation status of the certificates. It has another advantage due to the online and timely inquiry feature of the OCSP. However, the OCSP server's duties in the SMEmail differ from what is specified in RFC 2560. The OCSP server in the SMEmail should also verify the public key of the corresponding user using his/her certificate, if the queried certificate has a "good" status. The result of such verification should be included in the OCSP response. The OCSP responses are digitally signed with a private key that its corresponding trusted public key is known to the participants. Basic configuration of the SMEmail protocol is depicted in Fig. 3 in which $OCSP_A$ and $OCSP_B$ are the OCSP tokens for the certificates of *Alice* and *Bob* where *Alice* is the sender or initiator of the protocol and *Bob* is the designated recipient. The message exchange for basic configuration of the SMEmail is as follows.

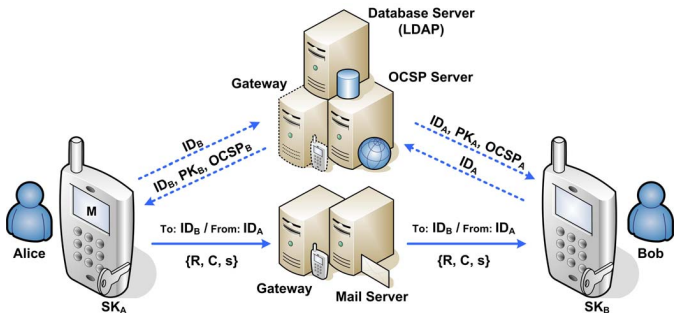

Figure 3. Basic configuration of the SMEmail

**SMEmail Composing:** The SMEmail application on the *Alice* side queries the OCSP server via the *Bob*'s identifier ($ID_B$) for his public key and the revocation status of his certificate. The OCSP server produces an OCSP response, digitally signs it, and sends ($ID_B$, $PK_B$, $OCSP_B$) to *Alice*. The SMEmail application on the *Alice* side verifies the OCSP server's signature, and uses $ID_B$ and $PK_B$ to generate $(R, C, s)$ from the *Alice*'s message $M$ by following the below steps:

(1) Randomly selects an integer $r \in_R [1, n-1]$.
(2) Computes $R = rG = (x_R, y_R)$.
(3) Computes $K = (r + \tilde{x}_R SK_A) PK_B = (x_K, y_K)$ where $\tilde{x}_R = 2^{\lceil f/2 \rceil} + (x_R \mod 2^{\lceil f/2 \rceil})$ in which $f = \lfloor \log_2 n \rfloor + 1$. If $K = O$ it goes back to step (1). Otherwise, it drives the secret key of encryption as $k = H'(x_K \| ID_A \| y_K \| ID_B)$ in which $H'$ is a one-way hash function that generates the required number of bits for the secret key of deployed symmetric encryption algorithm.
(4) Computes the ciphertext as $C = E_k(M)$ where $E(.)$ denotes the approved block cipher.
(5) Computes the digital signature as $s = tSK_A - r \pmod{n}$ where $t = H(M \| x_R \| ID_A \| y_R \| ID_B \| k)$.
(6) Composes a MIME entity containing $(R, C, s)$ as the message where the e-mail address of *Bob* is noted as the recipient, and forwards it to the mail server.

**SMEmail Delivering:** The SMEmail application on the *Bob* side queries the OCSP server via the *Alice*'s identifier ($ID_A$) for her public key and revocation status of her certificate. The OCSP server produces an OCSP response, digitally signs it, and sends ($ID_A$, $PK_A$, $OCSP_A$) to *Bob*. The SMEmail application on the *Bob* side verifies the OCSP server's signature and follows the below steps to extract the message of *Alice* and verify her signature:

(1) Computes $K = SK_B(R + \tilde{x}_R PK_A) = (x_K, y_K)$ and derives the secret key of encryption as $k = H'(x_K \| ID_A \| y_K \| ID_B)$.
(2) Decrypts the ciphertext as $M = D_k(C)$.
(3) Computes $t = H(M \| x_R \| ID_A \| y_R \| ID_B \| k)$.
(4) Verifies whether $sG + R \stackrel{?}{=} tPK_A$. If it is true, the extracted plaintext $M$ is accepted as the true message of *Alice*. Otherwise, it will be rejected.

The above-mentioned steps are regarded to the basic configuration of the SMEmail protocol that its basic configuration is depicted in Fig. 3. However, it is possible to enhance the performance by delegating the required validations to a trusted server. The required time of sending a certificate to a validation server, receiving, and authenticating the response can be considerably less than the required time of performing the certificate path discovery and validation on the resource constrained platforms such as mobile phones. The performance of SMEmail will be greatly enhanced by delegating the validations to a *Delegated Validation* (DV) server, as it is depicted in Fig. 4 and 5. The DV server accomplishes the certificate validation via the *Delegated Path Validation* (DPV) protocol [15]. However, its duties defer from what is specified in RFC 3379. The DV server accomplishes the certificate and public key validations for both *Alice* and *Bob*. It queries the database server for the certificates of the participants via their identifiers. It obtains the revocation statuses by getting OCSP responses from the OCSP server.

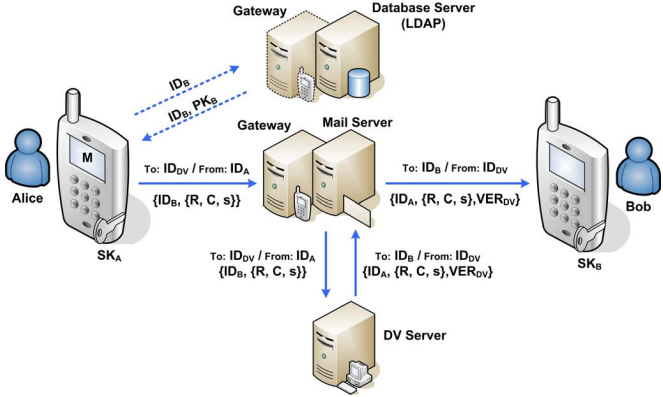

Figure 4. Delegated configuration of SMEmail using a typical Mail Server (the Mail Server is independent of the wireless network)

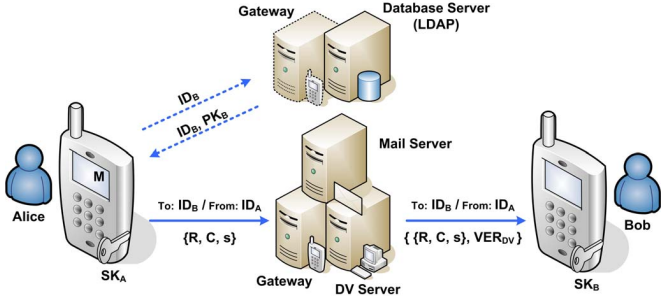

Figure 5. Delegated configuration of SMEmail when the Mail Server is adapted for the resource constraints (the Mail Server is provided by the wireless network or it is adapted for the mobile communications)

In the delegated configuration of SMEmail, all the secured e-mails are directed to the DV server. The DV server will contact the designated recipient after a successful validation. If any error occurs, the DV server sends an error message to the initiator and saves a copy in its log file. The DV server digitally signs its responses unless an error is occurred. The signed responses $VER_{DV}$ should include a hash value of all the transmitted parameters in addition to identifiers of both *Alice* and *Bob*. Here, *Bob* just checks the signature of DV server. *Alice* does not need to get any OCSP response for the certificate of *Bob*. She just needs to know the public key of *Bob* and save it in his phone for the future uses. If she does not know the public key of *Bob*, she can query it from the database server, as it is depicted in Fig. 4 and 5. Therefore, in the delegated configuration of SMEmail, the computational costs and communication overheads are efficiently decreased for the end-entities.

## IV. THE SECURITY OF SMEMAIL

The correctness of SMEmail can be simply verified. *Alice* and *Bob* reach to the same secret key of symmetric encryption as:

$$K_A = (r + \widetilde{x}_R SK_A)PK_B = (r + \widetilde{x}_R SK_A)SK_B G =$$
$$= SK_B(rG + \widetilde{x}_R SK_A G) = SK_B(R + \widetilde{x}_R PK_A) = K_B$$

The session key derivation function of SMEmail is an improved variant of the HMQV key establishment protocol [16]. However, it does not exactly correspond with the HMQV specifications. Defining $\widetilde{x}_R$ as the least significant half in binary representation of $x_R$ is just a trade-off between security and efficiency. Hereunder, some security attributes of the SMEmail are briefly described.

*1) Confidentiality*: The confidentiality is completely resided in the secrecy of session key since SMEmail uses a strong block cipher. The session key differs for different sessions and is derived from the private keys of the participants. The *Unknown Key-Share* (UKS) attack [17] is thwarted because the identifier of *Alice* is involved in derivation of the session key *k*. There are only two ways to defeat the confidentiality: finding $SK_B$, or having both $SK_A$ and *r*. Deriving the private keys and finding the corresponding *r* of a specific *R* are in deposit of solving the ECDLP that is computationally infeasible with the selected domain parameters of SMEmail.

*2) Authentication*: The implicit authentication is provided by SMEmail in three ways: SMEmail is a certificate-based protocol in which both of participants verify the certificate of the other party. An implicit authentication is also involved in the session key establishment so only the correct party who has the true private key can reach to the correct key agreement and perform the unsigncryption. Finally, the recipient verifies the signature by checking the $sG + R = tPK_A$ condition.

*3) Unforgeability*: It is computationally infeasible to forge a valid $(M, R, s)$ with $(M', R, s')$. A valid forged signature $s'$ should satisfy $s' = s + (t' - t)SK_A$ so the knowledge of *t*, *t'* and $SK_A$ is necessary. To find the true values of *t* and *t'*, the attacker should know the session key *k* that its knowledge is in deposit of his knowledge of $SK_B$, or both $SK_A$ and *r*. Otherwise, he cannot correctly forge the signature and his forged signature will be recognized in the delivering phase of SMEmail, when checking the $sG + R = tPK_A$ condition.

*4) Non-repudiation*: The non-repudiation can be deduced from the unforgeability. It is computationally infeasible to forge the signature of *Alice* without having her private key.

*5) Integrity*: The hash value of plaintext concatenated with some variable parameters is involved in the signature generation so any alteration in the plaintext will change the signature. The integrity is guaranteed by the security attributes of the deployed hash function and also unforgeability of the signature. The attacker should also have a valid session key to encrypt his modified message. Otherwise, his modified message will not be correctly decrypted by *Bob*. The session key is also involved in signature generation so he should have $SK_B$ or equivalently, both $SK_A$ and $r$ to establish a valid session key. He should also find a collision for the deployed hash function that is computationally infeasible. Otherwise, he cannot forge a valid signature. The integrity is implicitly verified when *Bob* checks the $sG + R = tPK_A$ condition.

6) *Forward secrecy of message confidentiality*: This means that even if the private key of *Alice* is revealed, the adversary is not capable of decrypting the previously signcrypted contents. As it was noted, the only way to defeat the message confidentiality is to have $SK_B$, or both of $SK_A$ and $r$. When $SK_A$ is revealed, for decryption to be possible, it is necessary to have the value of random number $r$ for the corresponding session that is in deposit of solving the ECDLP which is computationally infeasible with the selected domain parameters. As a one-pass protocol, we cannot prospect SMEmail for the *Perfect Forward Secrecy* (PFS) since there is not any session-specific input from *Bob*. However, it provides the partial forward secrecy under intractability of the ECDLP.

7) *Public verifiability*: Given $(R,C,M,k,s)$, anybody can compute $t = H(M \| x_R \| ID_A \| y_R \| ID_B \| k)$ and verify the signature of *Alice* by checking the $sG + R = tPK_A$ condition and without any need for the private keys of the participants.

## V. CONCLUSIONS

A new public key-based protocol for the secure e-mail in mobile environments (SMEmail) is introduced in this paper. It is an application-layer protocol that simultaneously provides the security attributes of confidentiality, integrity, authentication, non-repudiation, and forward secrecy of message confidentiality for the electronic mails. It effectively combines encryption and digital signature, and uses public keys for a secure key establishment where the established session key is used for enciphering contents via a symmetric encryption. SMEmail has great computational advantages over the traditional public key solutions, and is so suitable for mobile environments that are dealing with the resource constraints since it deploys elliptic curves and the concept of signcryption, and encrypts data using a symmetric enciphering algorithm, while it provides the most feasible security services.